\documentstyle[11pt]{article}
\normalbaselines

\newtheorem{lem}{Lemma}[section]

\begin{document}
\bibliographystyle{unsrt}

\def\bea*{\begin{eqnarray*}}
\def\eea*{\end{eqnarray*}}
\def\ba{\begin{array}}
\def\ea{\end{array}}
\count1=1
\def\be{\ifnum \count1=0 $$ \else \begin{equation}\fi}
\def\ee{\ifnum\count1=0 $$ \else \end{equation}\fi}
\def\ele(#1){\ifnum\count1=0 \eqno({\bf #1}) $$ \else \label{#1}\end{equation}\fi}
\def\req(#1){\ifnum\count1=0 {\bf #1}\else \ref{#1}\fi}
\def\bea(#1){\ifnum \count1=0   $$ \begin{array}{#1}
\else \begin{equation} \begin{array}{#1} \fi}
\def\eea{\ifnum \count1=0 \end{array} $$
\else  \end{array}\end{equation}\fi}
\def\elea(#1){\ifnum \count1=0 \end{array}\label{#1}\eqno({\bf #1}) $$
\else\end{array}\label{#1}\end{equation}\fi}
\def\cit(#1){
\ifnum\count1=0 {\bf #1} \cite{#1} \else 
\cite{#1}\fi}
\def\bibit(#1){\ifnum\count1=0 \bibitem{#1} [#1    ] \else \bibitem{#1}\fi}
\def\ds{\displaystyle}
\def\hb{\hfill\break}
\def\comment#1{\hb {***** {\em #1} *****}\hb }

\newcommand{\TZ}{\hbox{\bf T}}
\newcommand{\MZ}{\hbox{\bf M}}
\newcommand{\ZZ}{\hbox{\bf Z}}
\newcommand{\NZ}{\hbox{\bf N}}
\newcommand{\RZ}{\hbox{\bf R}}
\newcommand{\CZ}{\,\hbox{\bf C}}
\newcommand{\PZ}{\hbox{\bf P}}
\newcommand{\QZ}{\hbox{\rm eight}}
\newcommand{\HZ}{\hbox{\bf H}}
\newcommand{\EZ}{\hbox{\bf E}}
\newcommand{\GZ}{\,\hbox{\bf G}}

\font\germ=eufm10
\def\goth#1{\hbox{\germ #1}}
\vbox{\vspace{38mm}}

\begin{center}
{\LARGE \bf On $Q$-operators of XXZ Spin Chain of Higher Spin}\\[10 mm] 
Shi-shyr Roan \\
{\it Institute of Mathematics \\
Academia Sinica \\  Taipei , Taiwan \\
(email: maroan@gate.sinica.edu.tw ) } \\[25mm]
\end{center}

\begin{abstract}
We provide two methods of producing the $Q$-operator of XXZ spin chain of higher spin, one for $N$th root-of-unity $q$ with odd $N$ and another for a general $q$, as the generalization of those known in the six-vertex model. In the root-of-unity case, we discuss the functional relations involving the constructed $Q$-operator for the symmetry study of the theory. The $Q$-operator of XXZ chain of higher spin for a generic $q$ is constructed by extending Baxter's argument in spin-$\frac{1}{2}$ case for the six-vertex $Q$-operator. 

\end{abstract}
\par \vspace{5mm} \noindent
{\it 2006 PACS}:  05.50.+q, 02.20.Uw, 75.10.Pq \par \noindent

\setcounter{section}{0}
\section{Introduction}
\setcounter{equation}{0}
The objective of this paper is to construct the $Q$-operator(s) of XXZ spin chain of higher spin. The $Q$-matrices were invented by Baxter to solve the eigenvalue and eigenvector problems of the eight-vertex model \cite{B71, B72, B73, Bax}. The first one, denoted by $Q_{72}$,  was introduced by Baxter in 1972 \cite{B72} on some special "root-of-unity" crossing parameter $\eta$, where he computed eigenvalues of the eight-vertex model by establishing a functional equation, called the $TQ$-relation, between the transfer $T$-matrix and auxiliary $Q_{72}$-matrix. One year later in the computation of eigenvectors of the transfer matrix, he used a different $Q$-matrix in \cite{B73}. Further later, another $Q$-operator was constructed in \cite{Bax} for the study of eight-vertex model at the general crossing parameter $\eta$. All those $Q$-operators satisfy the same $TQ$-relation, which implies the Bethe equation of the model. In recent years, the degeneracy of the spin-$\frac{1}{2}$ XXZ Hamiltonian with the extra $sl_2$-loop-algebra symmetry was found and extensively analysed in \cite{DFM, De05, FM00, FM001, FM01} when the (anisotropic) parameter $q$ is a root of unity. The root-of-unity symmetry about the degenerate eigenspace of the six-vertex model has been further extended to the eight-vertex model \cite{De01a, De01b, FM02} and the XXZ spin chain of higher spin \cite{NiD, R06F}. Furthermore, a set of (conjectural) functional relations in the root-of-unity eight-vertex model was proposed by Fabricius and McCoy in \cite{FM04}, as an analogy to functional relations known in the $N$-state chiral Potts model (CPM) \cite{BBP, BazS}, where the chiral Potts transfer matrix in theory of CPM corresponds to some proper $Q$-operator of eight-vertex model that encodes the root-of-unity symmetry properties. Subsequently the functional-relation aspect on the  symmetry of solvable models suggested by Fabricius-McCoy comparison led to the discovery of the Onsager-algebra symmetry in the superintegrable CPM, parallel to the $sl_2$-loop-algebra symmetry in six-vertex model \cite{R04, R05o, R05b}. On the other hand, one may regard the CPM as a "model" theory in studying the degeneracy of solvable lattice models due to the better understanding of Baxter's $Q$-operator in CPM (i.e. the chiral Potts transfer matrix) in the functional-relation setting \cite{AMP, B90, B93, MR}.  Along this line, the investigation was successfully carried out in the root-of-unity six- and eight-vertex models \cite{F06, R06Q, R06Q8}. However in the case of XXZ spin chain of higher spin at roots of unity,  the functional relation study has not been conducted so far 
albeit the $sl_2$-loop-algebra symmetry of the system was already shown in \cite{R06F} using the algebraic Bethe-ansatz method and representation theory.  Indeed the $Q$-operator of XXZ spin chain of higher spin, except in the spin-$\frac{1}{2}$ case,  has not been explored in literature to the best of the author's knowledge. 
In the present article, we provide two methods  of producing $Q$-operators of XXZ spin chain of higher spin, one for a root-of-unity $q$ and another for the generic $q$, where in the root-of-unity case, the order of $q$, denoted by $N$, will assume to be {\it odd}.  For the root-of-unity $q$, the $Q$-operator is constructed by the same mechanism of producing the $Q$-operator of the root-of-unity six-vertex model in \cite{R06Q}, a method imitating the construction of Baxter's $Q_{72}$-operator in \cite{B72}.  For the general $q$, we extend  the argument of producing the six-vertex $Q$-matrix in \cite{Bax}  to the general case of XXZ chain with higher spin, then obtain the corresponding $Q$-operator. 

This paper is organized as follows. In section \ref{sec:XXZ}, we briefly review some basic facts about the XXZ spin chain of higher spin, that are needed for later discussions.
In section \ref{sec:QNq}, we construct the $Q$-operator of the spin-$\frac{d-1}{2}$ XXZ chain at $N$th root-of-unity $q$ with $d \leq N$, for the functional-relation study incorporating with the root-of-unity symmetry found in \cite{R06F}.
Section \ref{sec.Qgen} contains another $Q$-operator construction of the XXZ spin chain of higher spin, but valid for a generic $q$. By techniques in \cite{Bax} of producing the $Q$-operator of six-vertex model, we obtain a $Q$-operator of XXZ chain of higher spin with an arbitrary value of $q$. We close in section \ref{sec.F} with some concluding remarks.

\section{The XXZ Spin Chain of Higher Spin \label{sec:XXZ}}
\setcounter{equation}{0}
We first introduce here some basic concepts about the XXZ spin chain of higher spin needed for later discussions. The summary will be sketchy, also serves to establish notations, (for more details, see e.g. \cite{KiR, R06F, TakF} and references therein).

In this paper, the XXZ chain of spin $\frac{d-1}{2}$ for positive integers $d \geq 2$ will always assume with the {\it even} chain-size $L$.  We use ${\sf e}^k ~ ( 0 \leq k \leq d-1) $ to denote the standard basis of $\CZ^d$, and ${\sf e}_k ~ ( 0 \leq k \leq d-1) $ the dual basis. The local spin-operator of $\CZ^d$ will be denoted by ${\sf s}^z  := {\rm dia}[\frac{d-1}{2}, \frac{d-3}{2}, \ldots, \frac{-d+1}{2}]$, and $S^z$ is the total-spin operator:
$$
S^z = \sum_{\ell=1}^L {\sf s}_\ell^z .
$$
The $L$-operator of XXZ chain of spin-$\frac{d-1}{2}$ is the matrix with the $\CZ^2$-auxiliary and $\CZ^d$-quantum space: 
\be
{\sf L} (s)  =  \left( \begin{array}{cc}
        {\sf L}_{0,0}  & {\sf L}_{0,1}  \\
        {\sf L}_{1,0} & {\sf L}_{1,1} 
\end{array} \right)  ,  \ \  \ {\sf L}_{i,j} ~ (= {\sf L}_{i,j} (s) ) = \left( {{\sf L}_{i,j}}^{k'}_k \right)_{0 \leq k, k' \leq d-1} 
\ele(s6VL)
for $s \in \CZ$ with ${{\sf L}_{i,j}}^{k'}_k$  being zeros except 
\bea(lll)
{{\sf L}_{0,0}}^k_k &= {{\sf L}_{1,1}}^{d-k-1}_{d-k-1} &= a ( q^k s ) ~ ~ \ ~ \ ~ \ ~ \ ~ ~ ( 0 \leq k \leq d-1), \\ 
 {{\sf L}_{0,1}}^{d-k-2}_{d-k-1} &= 
{{\sf L}_{1,0}}^{k+1}_k &= q^{k+1}- q^{-k-1} ~ ~ (0 \leq k \leq d-2). 
\elea(Lij)
Hereafter $a(s)$ will denote 
the ($q$-dependent) $s$-function
$$
a(s)= s q^{\frac{-1 }{2}} - s^{-1} q^{\frac{1 }{2}}. 
$$
It is well-known that the $\CZ^d$-operators ${\sf L}_{i,j}$ give rise to the $d$-dimensional irreducible representation of $U_q(sl_2)$ (see, e.g. \cite{KiR, KRS}),
$$
\begin{array}{l}
{\sf L}_{0,0} = sq^{\frac{d-2}{2}} \widehat{K}^{\frac{-1}{2}} - s^{-1}q^{\frac{-(d-2)}{2}} \widehat{K}^{\frac{1}{2}} , ~ \ ~ \ ~ {\sf L}_{0,1} = (q-q^{-1}) \hat{e}_-, \\ 
{\sf L}_{1,0} = (q-q^{-1}) \hat{e}_+ , ~ \ ~ \ ~ {\sf L}_{1,1} = sq^{\frac{d-2}{2}}\widehat{K}^{\frac{1}{2}} - s^{-1}q^{\frac{-(d-2)}{2}} \widehat{K}^{\frac{-1}{2}} ,
\end{array}
$$
with $\widehat{K}^{\frac{1}{2}} = q^{S^z}$and $\widehat{K} \hat{e}_\pm \widehat{K}^{-1} = q^{\pm 2} \hat{e}_\pm$, $[\hat{e}_+, \hat{e}_- ] = \frac{\widehat{K} - \widehat{K}^{-1}}{q-q^{-1}}$.
The $L$-operator (\req(s6VL)) satisfies the YB relation
\be
R (\frac{s}{s'}) ({\sf L}(s) \bigotimes_{\rm aux}1) ( 1
\bigotimes_{\rm aux} {\sf L}(s')) = (1
\bigotimes_{\rm aux} {\sf L}(s'))( {\sf L}(s)
\bigotimes_{\rm aux} 1) R (\frac{s}{s'})
\ele(6YB)
with the $R$-matrix 
$$
R(s) = \left( \begin{array}{cccc}
        s^{-1} q - s q^{-1}  & 0 & 0 & 0 \\
        0 &s^{-1} - s & q-q^{-1} &  0 \\ 
        0 & q-q^{-1} &s^{-1} - s & 0 \\
     0 & 0 &0 & s^{-1} q - s q^{-1} 
\end{array} \right).
$$
The monodromy matrix of size $L$,
\be
\bigotimes_{\ell=1}^L {\sf L}_\ell (s) = \left( \begin{array}{cc}
        A(s)  & B(s) \\
        C(s) &  D(s) 
\end{array} \right), \ \ {\sf L}_\ell (t):= {\sf L}(t) \ {\rm at \ site} \ \ell, 
\ele(mM6d)
again satisfies the YB relation (\req(6YB)), hence 
its trace, called the transfer matrix of the XXZ chain of spin-$\frac{d-1}{2}$, forms a commutative family of $(\stackrel{L}{\otimes}\CZ^d)$-operators: 
\be
{\sf t}(s) = A(s) + D(s). 
\ele(sft)
Then $[{\sf t}(s),q^{S^z}]=0$ where $q^{S^z}=  \stackrel{L}{\otimes} \widehat{K}^{\frac{1}{2}}$. As the $R$-matrix at $s=q$ is of rank 1, the quantum determinant of (\req(mM6d)) is defined by 
$$
\begin{array}{c}
R(q) (\otimes {\sf L}_\ell (q s) \bigotimes_{\rm aux}1) ( 1
\bigotimes_{\rm aux} \otimes {\sf L}_\ell(s)) = \\
(1 \bigotimes_{\rm aux} \otimes {\sf L}_\ell (s))( \otimes {\sf L}_\ell (qs)
\bigotimes_{\rm aux} 1) R (q) 
=: {\rm det}_q \otimes {\sf L}_\ell (s) \cdot  R(q),
\end{array}
$$
or equivalently, 
\bea(c)
A(qs)C (s) =  C (qs) A (s) ,   B (qs) D (s)  =  D (qs)  B (s) , \\
B (s)  A(qs) =  A(s)  B(qs) ,  C(s) D(qs)  = D(s) C(qs) ; \\
{\rm det}_q \otimes {\sf L}_\ell(s) 
=A(qs) D(s) - C(qs) B(s) = D(qs)A(s) - B(qs)C(s) \\
~ \ ~ \ ~ \ ~ \ ~ \ ~ \ ~ \ ~ = D(s) A(qs) - C(s)B(qs) = A(s)D(qs)  - B(s)C(qs).
\elea(6qdf)
For the local $L$-operator (\req(s6VL)), ${\rm det}_q {\sf L}(s) = a( s) a(q^d s)$, so the quantum determinant of (\req(mM6d)) is equal to $ a(s)^L a(q^d s)^L$. 

\section{The $Q$-operator of XXZ Spin Chain of Higher Spin at Roots of Unity $q$ \label{sec:QNq}}
\setcounter{equation}{0}
In this section we study the $Q$-operator of the XXZ chain of spin $\frac{d-1}{2}$ at the $N$th root of unity $q$ with {\it odd} $N$ for $2 \leq d \leq N$. We will take $q$ and $q^{\frac{1}{2}}$ to be primitive $N$th roots of unity.  

As in the case of root-of-unity six-vertex model in \cite{R06Q}, the $Q_R$-matrix of the spin-$\frac{d-1}{2}$ XXZ chain is constructed from an ${\sf S }$-operator, which is a matrix with $\CZ^N$-auxiliary and $\CZ^d$-quantum space, ${\sf S }  = ({\sf S }_{i,j})_{i, j \in \ZZ_N} $ ~ $({\sf S }_{i,j}= {\sf S }_{i,j}(s))$, where the $\CZ^N$-auxiliary space has the basis indexed by  $\ZZ_N (:= \ZZ/N\ZZ)$, and ${\sf S }_{i,j}$ are $\CZ^d$-operators. The $Q_R$-operator is defined by 
$$
Q_R= {\rm tr}_{\CZ^N} ( \bigotimes_{\ell =1}^L {\sf S}_{\ell}), \ \ \ {\sf S}_{\ell}= {\sf S} \ {\rm at \ site} \ \ell.
$$
Then ${\sf t}Q_R = {\rm tr}_{\CZ^2 \otimes \CZ^N} ( \bigotimes_{\ell =1}^L {\sf U}_{\ell})$, where ${\sf U}_{\ell}= {\sf U}$ at the site $\ell$, and the local-operator ${\sf U}$ is the matrix with the $\CZ^2 \otimes \CZ^N $-auxiliary and $\CZ^d$-quantum space:
$$
{\sf U} = \left( \begin{array}{cc}
        {\sf L}_{0,0} {\sf S } & {\sf L}_{0,1} {\sf S} \\
        {\sf L}_{1,0} {\sf S} & {\sf L}_{1,1} {\sf S }
\end{array} \right) .
$$
The operator ${\sf t}Q_R$ will decompose into the sum of two matrices if we can find a $2N$ by $2N$ scalar matrix ${\sf M}$ (independent of $s$) of the form 
\be
{\sf M} = \left( \begin{array}{cc}
        I_N  & 0 \\
        \delta & I_N 
\end{array} \right) , \ \ \delta = {\rm dia} [\delta_0, \cdots, \delta_{N-1}] .
\ele(Md)
such that $
{\sf M}^{-1} {\sf U} {\sf M} = \left( \begin{array}{cc}
        * & * \\ 
         0 & *
\end{array} \right)$. One can express ${\sf M}^{-1} {\sf U} {\sf M}$ by
$$
{\sf M}^{-1} {\sf U} {\sf M} = \left( \begin{array}{cc}
        {\sf A}(\delta_j){\sf S }_{i,j}  ,  & {\sf L}_{0,1} {\sf S }_{i,j}   \\ 
       {\sf C}(\delta_i, \delta_j){\sf S }_{i,j}  , & {\sf D}(\delta_i) {\sf S }_{i,j} 
\end{array} \right)_{i, j \in \ZZ_N} , 
$$
where ${\sf A}(\delta_j), {\sf C}(\delta_i, \delta_j), {\sf D}(\delta_i)$ are the $\CZ^d$-operators
\bea(l)
{\sf A}(\delta_j)(s):= {\sf L}_{0,0}  +  {\sf L}_{0,1} \delta_j,  ~ ~ ~  {\sf D}(\delta_i)(s):= - \delta_i {\sf L}_{0,1}   + {\sf L}_{1,1} ,   \\
{\sf C}(\delta_i, \delta_j) (s) := - \delta_i {\sf L}_{0,0}  + {\sf L}_{1,0}  - \delta_i  {\sf L}_{0,1} \delta_j  +{\sf L}_{1,1} \delta_j .
\elea(gauL) 
The above operators satisfies the following commutative relations for arbitrary $\delta_i, \delta_j$:
\bea(l)
{\sf C}(\delta_i, \delta_j) (qs) {\sf A}(\delta_j)(s) =  {\sf A}(\delta_j) (qs) {\sf C}(\delta_i, \delta_j)(s), \\
{\sf C}(\delta_i, \delta_j) (s) {\sf D}(\delta_i)(qs) =  {\sf D}(\delta_i) (s) {\sf C}(\delta_i, \delta_j)(qs) .
\elea(ADCg)
In fact, by the quantum determinant property (\req(6qdf)) of the $L$-operator, one finds 
$$
\begin{array}{l}
({\sf L}_{1,0} + {\sf L}_{1,1} \delta_j)(qs){\sf A}(\delta_j)(s) = {\sf A}(\delta_j)(qs)({\sf L}_{1,0}+ {\sf L}_{1,1} \delta_j)(s), \\
(- \delta_i {\sf L}_{0,0}  + {\sf L}_{1,0})(s) {\sf D}(\delta_i)(qs) = {\sf D}(\delta_i)( s) (- \delta_i {\sf L}_{0,0}  + {\sf L}_{1,0})(qs) .
\end{array}
$$
Then the equality, $
{\sf C}(\delta_i, \delta_j) = - \delta_i {\sf A}(\delta_j) + {\sf L}_{1,0} + {\sf L}_{1,1} \delta_j = - \delta_i {\sf L}_{0,0}  + {\sf L}_{1,0}  +{\sf D}(\delta_i) \delta_j$,  
in turn yields the relation (\req(ADCg)).

\begin{lem}\label{lem:ADv}
Let ${\sf A}(\beta), {\sf D}(\alpha)$ and ${\sf C}(\alpha, \beta)$ be the operators defined in $(\req(gauL))$ for $\alpha, \beta \in \CZ^*$ and $q$ an arbitrary number (not necessary a root of unity). Then the criterion of $\alpha, \beta$ with the zero determinant function, ${\rm det}{\sf C}(\alpha, \beta) (s)=0$,  is: $\beta = q^{d-1- 2k} \alpha$ for some $ 0 \leq k \leq d-1$. In this situation, ${\sf C}(\alpha, \beta)(s)$ has one-dimensional kernel and cokernel for all $s \in \CZ$, generated by a non-zero vector ${\sf v} ( = {\sf v} (s)) \in \CZ^d$, $\widehat{\sf v} ( = \widehat{\sf v} (s)) \in \CZ^{d*}$ (unique up to non-zero multiples of scalar $s$-functions) with  ${\sf C}(\alpha, \beta)(s) {\sf v }= \widehat{\sf v }{\sf C}(\alpha, \beta)(s)= 0$. Furthermore the following relations hold: 
\bea(l)
{\sf A}(\beta)(s){\sf v }(s) = a(s) \frac{{v(s)}_0}{{v (qs)}_0} {\sf v }(qs), ~ \ ~ \ ~ ~ {\sf D}(\alpha)(s){\sf v }(s) = a(q^{d-1}s) \frac{{v(s)}_0}{{v(\frac{s}{q})}_0} {\sf v}(\frac{s}{q}), \\
\widehat{\sf v}(s) {\sf A}(\beta) (s) = a(q^{d-1}s) \frac{{\widehat{v}(s)}^{d-1}}{{\widehat{v}(\frac{s}{q})}^{d-1}} \widehat{\sf v}(\frac{s}{q}), ~
 \widehat{\sf v} (s) {\sf D}(\alpha) (s)  = a(s)\frac{{\widehat{v}(s)}^{d-1}}{{\widehat{v}(qs)}^{d-1}} \widehat{\sf v}(qs) ,  
\elea(vADv)
where ${{\sf v}(s)}_0 = {\sf e}_0({\sf v}(s))$, ${\widehat{\sf v}(s)}^{d-1} = \widehat{\sf v}(s)({\sf e}^{d-1})$, the ${\sf e}^0$- and ${\sf e}_{d-1}$-component of $v(s)$, $\widehat{v}(s)$, respectively.
\end{lem} 
{\it Proof.} The relation $\beta = q^{d-1- 2k} \alpha$ with $ 0 \leq k \leq d-1$ follows from the equality 
$$
{\rm det}{\sf C}(\alpha, \beta) (s) = {\rm det}( \beta - \alpha q^{2S^z} ) \prod_{k=0}^{d-1} a(q^k s).
$$
By (\req(Lij)), the $\CZ^d$-operator ${\sf C}(\alpha, \beta) (s)$ in (\req(gauL)) has the rank at least $d-1$ in general. When $\beta = q^{d-1- 2k} \alpha$, it has the one-dimensional kernel and cokernel for all $s \in \CZ$, hence generated by the non-zero vector ${\sf v}(s) \in \CZ^d$, $\widehat{\sf v}(s) \in \CZ^{d*}$, respectively. In this situation, by (\req(ADCg)), $
{\sf C}(\alpha, \beta) (qs) ({\sf A}(\beta)(s){\sf v}(s))  =
{\sf C}(\alpha, \beta) (q^{-1}s) ({\sf D}(\alpha)(s){\sf v}(s)) = 0$. Since ${\sf v}(qs), {\sf v}(q^{-1} s)$ can be characterized as the solution, unique up to scalar $s$-functions, for the equation 
$${\sf C}(\alpha, \beta) (qs) {\sf v}(qs) = {\sf C}(\alpha, \beta) (q^{-1}s){\sf v }(q^{-1} s) =0,$$
the expressions of ${\sf A}(\beta), {\sf D}(\alpha)$ in turn yield the first two relations in (\req(ADSij)) about ${\sf v}$. Similarly, follow the other two relations in (\req(ADSij)) about $\widehat{\sf v}$. 
$\Box$ 
\par \vspace{.1in} \noindent
{\bf Remark}. The closed form of the vector ${\sf v}$ and covector $\widehat{\sf v}$ in the above lemma is in general hard to obtain except the cases $\beta = q^{\pm (d-1)} \alpha$, where ${\sf v }= \sum_{k=0}^{d-1} {\sf v }_k {\sf e}^k$, $\widehat{\sf v} = \sum_{k=0}^{d-1} {\sf e}_k \widehat{\sf v}^k $ are given by\footnote{For $d=2$, the ${\sf v}(s), {\sf v}(qs)$ in (\req(vform)) here correspond to the $g_i, g'_i$ in \cite{Bax} (9.8.17).}
\be
\frac{{\sf v }_k}{{\sf v }_0} = \beta^k (- s q^{\frac{-1}{2}})^{\pm k} \left[ {}^{d-1}_{~ k ~ }  \right]_q , \ ~ \ 
\frac{\widehat{\sf v}^k}{\widehat{\sf v}^0} = \alpha^{-k} (s q^{\frac{-1}{2}})^{\pm k}.
\ele(vform)
Here $ \left[ {}^m_n   \right]_q  = \frac{[m] !}{[m-n] ! [n] !}$ is 
the  $q$-binomial for integers $0 \leq n \leq m$ with $[n] = \frac{q^n- q^{-n}}{q-q^{-1}}$, $[n]! = \prod_{i=1}^n  [i]$ and $[0]!:=1$. Indeed in the above cases, one can directly verify the relation (\req(vADv)) using the formula (\req(vform)).

In the procedure of constructing $Q_R$-operator in the $N$th root of unity $q$,  
the necessary condition for $\delta_i, \delta_j$ with the zero lower blocktriangular matrix of ${\sf M}^{-1} {\sf U} {\sf M}$, $
{\sf C}(\delta_i, \delta_j)(s) {\sf S }_{i,j}  = 0 $ for ${\sf S}_{i,j} \neq 0$, 
is the vanishing determinant of ${\sf C}(\delta_i, \delta_j)(s)$ for all $s$. By Lemma \ref{lem:ADv}, $\delta_j = q^{d-1- 2k} \delta_i$ for some $ 0 \leq k \leq d-1$, and 
the non-zero ${\sf S }_{i,j}$ with ${\sf C}(\delta_i, \delta_j)(s) {\sf S }_{i,j} (s)  = 0$ is expressed by  
\be
{\sf S }_{i,j} ( = {\sf S }_{i,j}(s)) = {\sf v}_{i,j}(s) \tau_{i, j}, ~  ~ {\sf v}_{i,j}(s) \in \CZ^d ,  ~ \tau_{i, j} \in \CZ^{d *}
\ele(CS0)
where ${\sf v}_{i,j} (={\sf v}_{i,j}(s)) $ is the kernel vector of ${\sf C}(\delta_i, \delta_j)(s)$.
Hereafter we shall always choose the above dual-vector $\tau_{i, j}$ as parameters independent of $s$. By (\req(vADv)), we find
\bea(l)
{\sf A}(\delta_j)(s){\sf S }_{i,j}(s) = a(s) \frac{{v_{i,j}(s)}_0}{{v_{i,j}(qs)}_0} {\sf S }_{i,j}(qs), \\ {\sf D}(\delta_i)(s){\sf S }_{i,j}(s) = a(q^{d-1}s) \frac{{v_{i,j}(s)}_0}{{v_{i,j}(s/q)}_0} {\sf S }_{i,j}(\frac{s}{q}),
\elea(ADSij)
where ${{\sf v}_{i,j}(s)}_0 = {\sf e}_0({\sf v}_{i,j}(s))$. For the $Q_L$-operator, we repeat the above working, replacing ${\sf S}$,  ${\sf L}(s){\sf S}$ by $\widehat{\sf S}$, $\widehat{\sf S}{\sf L}(s)$ with the same ${\sf M}$. We form the product $Q_L{\sf t}$ using $Q_L= {\rm tr}_{\CZ^N} ( \bigotimes_{\ell =1}^L \widehat{\sf S}_{\ell})$, and  $\widehat{\sf S}= (\widehat{\sf S}_{i,j})_{i,j \in \ZZ_N}$ with 
\be
\widehat{\sf S}_{i, j} (= \widehat{\sf S}_{i, j}(s)) = \widehat{\tau}_{i, j}\widehat{\sf v}_{i, j}(s), \ \ \ \ \ \widehat{\tau}_{i, j} \in \CZ^d , \ \  \widehat{\sf v}_{i, j}(s) \in \CZ^{d*},
\ele(hSij)
where $\widehat{\sf v}_{i,j} (=\widehat{\sf v}_{i,j}(s)) $ is the cokernel vector of ${\sf C}(\delta_i, \delta_j)(s)$. By (\req(vADv)), follow the relations 
$$
\begin{array}{l}
\widehat{\sf S}_{i, j}(s) {\sf A}(\delta_j) (s) = a(q^{d-1}s) \frac{{\widehat{v}_{i,j}(s)}^{d-1}}{{\widehat{v}_{i,j}(s/q)}^{d-1}} \widehat{\sf S}_{i, j}(\frac{s}{q}), \\
 \widehat{\sf S}_{i, j}(s) {\sf D}(\delta_i) (s)  = a(s)\frac{{\widehat{v}_{i,j}(s)}^{d-1}}{{\widehat{v}_{i,j}(qs)}^{d-1}} \widehat{\sf S}_{i, j}(qs) 
\end{array}
$$
where ${\widehat{\sf v}_{i,j}(s)}^{d-1} = \widehat{\sf v}_{i,j}(s)({\sf e}^{d-1})$.

We now consider the case of $\delta_i, \delta_j$ with the relation $\delta_j = q^{\pm (d-1)} \delta_i$. 
Choose the diagonal matrix $M$ in (\req(Md)) as 
\be
\delta = {\rm dia}[1, q^{d-1}, \cdots, q^{(N-1)(d-1)}], 
\ele(diaQ)
i.e. $\delta_i = q^{(d-1)i}$ for  $ 0 \leq i \leq N-1$. 
As the transfer matrix ${\sf t}(s)$ (of even chain-size) is the same when changing the $L$-operator ${\sf L}(s)$ in (\req(s6VL)) to ${\sf L} (-s)$, we shall use ${\sf S}_{i, j}= {\sf S}_{i, j}(-s)$ in (\req(CS0)) for the construction of $Q_R$-operator, (i.e. replacing $a(s)$ by $a(-s)$, and $s$ by $-s$ for the vector ${\sf v}_{i,j}$ in (\req(vform))), but keep $\widehat{\sf S}_{i, j} = \widehat{\sf S}_{i, j}(s)$ in (\req(hSij)) for the $Q_L$-operator. By setting $(v_{i,j})_0= \delta_j^{\frac{-(d-1)}{2}}(s q^{\frac{-1}{2}})^{\frac{\mp (d-1)}{2}} $,  $(\widehat{v}_{i,j})^0= \delta_i^{\frac{d-1}{2}} (s q^{\frac{-1}{2}})^{\frac{\mp (d-1)}{2}} $ for vectors in (\req(vform)), ${\sf S}_{i, j}, \widehat{\sf S}_{i, j}$ are defined to be zero except $i-j = \pm 1 \in \ZZ_N$, in which cases the ${\sf v}_{i, j}, \widehat{\sf v}_{i, j}$ in (\req(CS0)), (\req(hSij)) are given by\footnote{Note that for $d=2$, vectors in (\req(vhvij)) here are the same as \cite{R06Q} (4.4) and (4.7).}
\bea(l)
{\sf v}_{i, j} (= {\sf v}_{i, j}  (s)) = 
 \| q^{(d-1)j}(sq^{\frac{-1}{2}})^{j-i} \rangle   , \\
\widehat{\sf v}_{i, j} (= \widehat{\sf v}_{i, j}(s) )   =  \langle q^{-(d-1)i} (s q^{\frac{-1}{2}} )^{j-i} \| . 
\elea(vhvij)
Here the vector $\| s \rangle$ and covector $\langle s \|$ for $s \in \CZ^*$ are defined by
$$
\| s \rangle = \sum_{k=0}^{d-1} s^{ \frac{-(d-1)}{2} + k } \left[ {}^{d-1}_{~ k ~ }  \right]_q  {\sf e}^k \in \CZ^d , \ ~ \ ~ \langle s \| = \sum_{k=0}^{d-1} s^{ \frac{-(d-1)}{2} + k }  {\sf e}_k \in \CZ^{d *}.
$$
The ${\sf S}$-operator is now in the form 
$$
{\sf S} = \left( \begin{array}{cccccc}
        0  & {\sf S}_{0,1} & 0&\cdots   & 0&{\sf S}_{0,N-1}\\
        {\sf S}_{1,0} &  0& {\sf S}_{1,2} & \ddots & &\vdots \\
0&&\ddots&\ddots&& \\
\vdots&&\ddots&\ddots&&0 \\
0&&&\ddots& &{\sf S}_{N-2,N-1}\\
{\sf S}_{N-1,0} &0&\cdots &0&{\sf S}_{N-1,N-2}&0
\end{array} \right) ,
$$
and the same for the $\widehat{\sf S}$-operator.
By (\req(ADSij)), one finds
$$
\begin{array}{l}
{\sf A}(\delta_j)(-s){\sf S }_{i,j}(s) = a(-s)q^{\frac{(j-i)(d-1)}{2}} {\sf S }_{i,j}(qs),  \\ {\sf D}(\delta_i)(-s){\sf S }_{i,j}(s) = a(-q^{d-1}s) q^{\frac{-(j-i)(d-1)}{2}}  {\sf S }_{i,j}(s/q),
\end{array}
$$
then by the even chain-size, follows  
\be
{\sf t}(s) Q_R (s) = a(q^{d-1}s)^L  Q_R ( s q^{-1}) + a(s)^L Q_R ( s q) .
\ele(TQR)
Similarly, one has 
\be
Q_L (s) {\sf t}(s)  = a(q^{d-1}s)^L  Q_L ( s q^{-1}) + a(s)^L Q_L ( sq) .
\ele(QLT)
In order to construct the $Q$-operator from $Q_R$ and $Q_L$, it suffices to show (\cite{B72} (C28)):
\be
Q_L (s') Q_R (s) = Q_L (s) Q_R (s') \ ~ \ {\rm for} \ s, s' \in \CZ^*.
\ele(QLR)
To do this,  as in \cite{B04, R06Q} we consider the product function of (\req(vhvij)) , $f(s', s | i, j ; k, l) := \widehat{\sf v}_{i, j}(s') {\sf v}_{k, l} (s)$, which is equal to zero except 
$|j-i| =|l-k|=1$, in which case $f(s', s | i, j ; k, l)$ is expressed by
$$
\sum_{r =0}^{d-1} q^{ \frac{(i-l)(d-1)((d-1)-2r)}{2}  } \left( (s'q^{\frac{-1}{2}})^{(j-i) } 
(sq^{\frac{-1}{2}})^{(l-k) }\right)^{\frac{-(d-1)+2r}{2}} \left[ {}^{d-1}_{~ r ~ }  \right]_q .
$$
Next look for an auxiliary function $P( s', s | n)$ for $ n \in \ZZ_N$ such that  
\be 
f(s, s' | i, j ; k, l) = P( s', s | k-i) f(s', s | i, j ; k, l) P( s', s | l-j)^{-1},
\ele(fP)
by which the product $Q_L (s) Q_R (s')$ differs by the boundary contribution when interchanging $s$ by $s'$,   
$$
Q_L (s) Q_R (s') = P(s', s | k_1-i_1) Q_L (s') Q_R (s) P(s', s | k_{L+1}-i_{L+1})^{-1} ,
$$
hence follows the relation (\req(QLR)) due to the periodicity of boundary condition. There are four cases, $|j-i|=|l-k|=1$, in (\req(fP)) to consider the function $P$, which automatically holds for $j-i= l-k$. The other two cases yield just one condition on the function $P$: $\frac{P(s', s | n+2)}{P(s', s | n)} = \frac{F(s/s'| n )}{F(s'/s | n )}$, where $F (s | n)$ are defined by 
$$
F (s | n) =  \sum_{r=0}^{d-1} q^{(n+1)(d-1)(\frac{-d+1}{2}+ r)} s^{\frac{-d+1}{2}+ r} \left[ {}^{d-1}_{~ r ~ }  \right]_q , \ \ ( s \in \CZ^*, n \in \ZZ_N). 
$$
Since $F(s |n)= F(s^{-1} | n')$ for $n+n'+2 \equiv 0 \pmod{N}$, there are functions $P(s', s | n)$ satisfying the relation (\req(fP)). Hence follows (\req(QLR)). 

We now define  
\be
Q(s) := Q_R(s) Q_R(s_0)^{-1} = Q_L (s_0)^{-1} Q_L(s) ,
\ele(Qdef)
where $s_0$ is a fixed value of $s$ so that $Q_R(s_0)$ and $Q_L(s_0)$ are non-singular. By (\req(TQR)), (\req(QLR)), one finds 
$[ Q (s), Q (s') ]= [ {\sf t} (s), Q (s') ] = 0$, and the $TQ$-relation
\be
{\sf t} (s) Q (s)   =  a(q^{d-1}s)^L  Q ( s q^{-1}) + a(s)^L Q ( s q)  .
\ele(stQ)
As  in the discussion of $Q$-operator of the eight-vertex model in \cite{B72, FM04},  the non-singular property of $Q_R(s_0), Q_L (s_0)$ for a general $s_0$ with generically arbitrary $\tau_{i, j}, \widehat{\tau}_{i, j}$ in (\req(CS0)), (\req(hSij)) is expected, (unfortunately I know of no simple way  to prove it). Nevertheless we shall assume the non-singular property of $Q_R(s_0), Q_L (s_0)$ for some $s_0$, and define the $Q$-operator in (\req(Qdef)), which is independent to the choice of $\tau_{i, j}, \widehat{\tau}_{i, j}$ regardless of $Q_R, Q_L$'s dependence  on them. Since vectors ${\sf v}_{i, j}, \widehat{\sf v}_{i, j}$ in (\req(vhvij)) satisfy the relation, $q^{-(d-1) {\sf s}^z  } {\sf v}_{i, j} (s) = {\sf v}_{i+1, j+1} (s) $, $\widehat{\sf v}_{i, j} (s) q^{(d-1) {\sf s}^z} = \widehat{\sf v}_{i+1, j+1}(s) $, the relations 
$$
q^{-(d-1) {\sf s}^z}{\sf S}_{i, j} q^{(d-1) {\sf s}^z} = {\sf S}_{i+1, j+1}, ~ 
q^{-(d-1) {\sf s}^z}\widehat{\sf S}_{i, j}q^{(d-1){\sf s}^z}= \widehat{\sf S}_{i+1, j+1},
$$ 
will hold if the parameters $\tau_{i, j}, \widehat{\tau}_{i, j}$ in (\req(CS0)), (\req(hSij)) are chosen with $\tau_{i, j}q^{(d-1){\sf s}^z}=\tau_{i+1, j+1} $, $q^{-(d-1){\sf s}^z} \widehat{\tau}_{i, j}=\widehat{\tau}_{i+1, j+1}$, (e.g., by setting 
$\tau_{i, j}=\widehat{\sf v}_{i, j}(c)$, $\widehat{\tau}_{i, j}={\sf v}_{i, j}(c)$
for some fixed complex number $c$). This in turn yields $Q_R, Q_L$, hence the $Q$-operator, commute with $q^{(d-1) S^z}$. When $d-1$ is relatively prime to $N$, e.g. the cases $d=2,3, N-1$, one has $[Q(s), q^{S^z}]=0$.

For $j \geq 2$, the $j$th fused $L$-operator for the spin-$\frac{d-1}{2}$ XXZ spin chain is the matrix $L^{(j)}(s) = \left( L^{(j)}_{k, l} (s) \right)_{0 \leq k, l \leq j-1}$ of $\CZ^j$-auxiliary and $\CZ^d$-quantum space constructed from (\req(s6VL)). The  
$\CZ^j$-auxiliary space is the space of completely symmetric $(j-1)$-tensors of $\CZ^2$ with the basis $e^{(j)}_k = \widehat{x}^{j-1-k} \widehat{y}^k$ and the dual basis $e^{(j) *}_k = {j-1-k \choose k} x^{j-1-k} y^k$ ~ $(k=0, \ldots, j-1)$, where $\widehat{x}, \widehat{y}$ denote the standard basis  of $\CZ^2$-auxiliary space in (\req(s6VL)),  $ x, y $ the dual basis, 
$\widehat{x}^m \widehat{y}^n$ the symmetric $(m+n)$-tensor of $\CZ^2$ defined by 
$$
{m+n \choose n} \widehat{x}^m \widehat{y}^n = \underbrace{\widehat{x}\otimes \ldots \otimes \widehat{x}}_{m} \otimes \underbrace{\widehat{y}\otimes \ldots \otimes \widehat{y}}_{n} + \ {\rm others \ by \ permutations}, 
$$
and the same for $x^m y^n$. The $\CZ^d$-operator $L^{(j)}_{k, l} (s)$ is  defined in \cite{R06F} with the expression\footnote{Note that $L^{(j)}_{k, l} (s)$ here differs from ${{\sf L}^{(j)}_{k, l} (s)}$ in \cite{R06F} by a factor, $L^{(j)}_{k, l} (s)= \frac{\prod_{i=1}^N a(q^i s)}{\prod_{i=d-1}^{N+j-2} a(q^i s) } {{\sf L}^{(j)}_{k, l} (s)}$, contributed from the normalized factor in the (corrected) formula (4.27) of \cite{R06F}: ${\sf T}^{(j)} (t) := \frac{\omega^{-(j-1)S^z} (1-t^N)^L {\cal T}^{(j)}(t)}{\prod_{k=d-2}^{N+j-3} h(\omega^k t)^L }, j \geq 1$. Hence the fusion matrix $T^{(j)} (s)$ in (\req(Tj)) of this paper and ${\sf t}^{(j)} (s)$ in \cite{R06F} are related by $T^{(j)} (s)= \frac{\prod_{i=1}^N a(q^i s)^L}{\prod_{i=d-1}^{N+j-2} a(q^i s)^L } {\sf t}^{(j)} (s)$.  }
\begin{eqnarray*}
L^{(j)}_{k, l} (s) = \frac{\prod_{i=1}^N a(q^i s)}{\prod_{i=d-1}^{N+j-2} a(q^i s) }  \langle e^{(j)*}_k | {\sf L}(s) \otimes_{\rm aux} {\sf L}(qs)\otimes_{\rm aux} \cdots \otimes_{\rm aux} {\sf L}(q^{j-2}s) | e^{(j)}_l \rangle .
\end{eqnarray*}
The $j$th fusion matrix is the trace of the monodromy matrix
\be
T^{(j)} (s) = {\rm tr}_{\CZ^j} (\bigotimes_{\ell=1}^L  L^{(j)}_\ell (s)), \ ~  \ L^{(j)}_\ell (s) = L^{(j)}(s) \ {\rm at ~ site} ~ \ell ,
\ele(Tj)
which form a family of commuting operators of $\stackrel{L}{\otimes} \CZ^d$. Since $
L^{(2)} (s) =  \prod_{i=1}^{d-2} a(q^i s)  {\sf L}(s)$, we have 
$T^{(2)} (s)=  \prod_{i=1}^{d-2} a(q^i s)^L  {\sf t} (s)$, hence the $TQ$-relation by (\req(stQ)): 
\be
 T^{(2)} (s) Q (s)   =   h(s)^L  Q ( q^{-1}s) +  h( q^{-1} s)^L Q ( qs )  
\ele(TQ)
where the function $h(s)$ is defined by
$$
h (s) ( = h_d (s) ) = \prod_{i=1}^{d-1} a(q^i s).
$$
Using \cite{R06F} (4.18), one finds the fusion relations among $T^{(j)}$'s by setting $T^{(0)}=0, T^{(1)}=h(q^{-1}s)^L$:
\be 
T^{(j)}(s) T^{(2)}(q^{j-1}s) = h (q^{j-1}s)^L T^{(j-1)}(s) + h (q^{j-2}s)^L  T^{(j+1)}(s) 
\ele(Fus)
for $j \geq 1$.
By the induction argument, the $TQ$- and fusion relations, (\req(TQ)) , (\req(Fus)), in turn yield the $T^{(j)}Q$-relation 
\be
T^{(j)}(s) = Q( q^{-1} s )Q( q^{j-1}s )\sum_{k=0}^{j-1} \frac{h (q^{k-1} s)^L}{ Q(q^{k-1}s) Q(q^k s)} 
\ele(TjQ)
for $j \geq 1$.
Then follows the boundary fusion relation:
\be 
T^{(N+1)}(s) =  T^{(N-1)}(q s) + 2  h (q^{-1} s)^L .
\ele(bFu)

Similar to the functional equation satisfied by the chiral Potts transfer matrix (\cite{BBP} (4.40)) in CPM , the $Q$-operator of  XXZ spin chain of higher spin encoding the root-of-unity property is expected to satisfy the $Q$-functional equation as in the cases of six-vertex and eight-vertex models \cite{FM04, FM41, R06Q, R06Q8}:  
\be
Q ( C s ) = M_0 Q ( s ) \sum_{k=0}^{N-1} \frac{ h (q^k s)^L}{ Q (q^k s) Q (q^{k+1} s) }  
\ele(Qeq)
where the $s$-automorphism $C$ is conjecturally defined by $C(s )= -s$, and $M_0$ is some normalized matrix.  By (\req(TjQ)), the $Q$-functional equation (\req(Qeq)) is equivalent to the $N$th $QQ$-relation 
\be
 T^{(N)}(q v) = M^{-1}_0 Q ( C s ) Q(s)  , 
\ele(QQN)
or either one of the following $QQ$-relations:  
\be
T^{(j)}(q s) + T^{(N-j)}(q^{j+1} s) = M_0^{-1} Q ( C q^j s )Q(s) 
\ele(QQ)
for $0 \leq j \leq N$.
Note that from the structure of eigenvalues of the $Q$-operator, the $TQ$-relation (\req(TQ)) in turn yields the well-known Bethe equation of spin-$\frac{d-1}{2}$ XXZ chain (see, e.g. \cite{KiR, R06F} and references therein):
$$
 \frac{ a(s_i )^L }{a(q^{d-1} s_i )^L}  = -  q^{-2r} \prod_{k =1}^m \frac{s^2_k -q^{-2} s^2_i }{  s^2_k - q^2  s^2_i  }, \ ~ ~ \ ~ i =1, \ldots, m ,
$$
with the evaluation parameters  of the $sl_2$-loop-algebra symmetry  (see \cite{R06F} (4.32)) built into the $QQ$-relation (\req(QQN)).

\section{ The Matrix $Q(s)$ of XXZ Spin Chain of Higher Spin for a General $q$ \label{sec.Qgen}}
\setcounter{equation}{0}
We now study the XXZ spin chain of higher spin with an arbitrary value of $q$ (not necessary a root of unity), which we will assume in this section.   
By imitating Baxter's method of producing the $Q$-operator of six-vertex model in \cite{Bax} section 9.8., we construct a $Q$-operator of XXZ spin chain of spin $\frac{d-1}{2} ~ (d \geq 2)$ for a general $q$. 

We search vectors as columns of $Q_R(s)$, whose image under ${\sf t}(s)$, by (\req(TQR)), is  the sum of two proportional vectors with twisted variables. To do this, we perform the gauge transform of the monodromy matrix (\req(mM6d)) by choosing two-by-two matrices $P_1, \ldots, P_L$ and $P_{L+1} = P_1$ of the form
$$
P_\ell = \left( \begin{array}{cc}
        1  & 0 \\
        r_\ell &  1 
\end{array} \right) \ ~ \ \ ( 1 \leq \ell \leq L+1) , \ \ r_{L+1}= r_1 , 
$$
such that the bottom-left entry of $P_\ell^{-1} {\sf L}_\ell (s) P_{\ell+1}$ has a non-trivial kernel vector in $\CZ^d$ for $1 \leq \ell \leq L$.  Under the gauge transform by $P_\ell$'s, the $L$-operator (\req(s6VL)) at site $\ell$ of the monodromy matrix is replaced by
$$
P_\ell^{-1} {\sf L}(s) P_{\ell+1} = \left( \begin{array}{cc}
        {\sf A}(r_{\ell+1})(s)  &  {\sf L}_{0,1}(s) \\
       {\sf C}(r_\ell, r_{\ell+1})(s),  & {\sf D}( r_\ell ) (s)
\end{array} \right)
$$
where ${\sf A}(r_{\ell+1})$, ${\sf D}( r_\ell ) $ and ${\sf C}(r_\ell, r_{\ell+1})$ are the operators defined in (\req(gauL)). By Lemma \ref{lem:ADv}, the condition of ${\sf C}(r_\ell, r_{\ell+1})$ with the non-trivial kernel is given by $r_{\ell+1} = r_\ell q^{d-1 - 2k}$ for some $0 \leq k \leq d-1$. Thus the general solution for $r_\ell$'s is : $
r_\ell = r q^{\varepsilon_1 + \ldots + \varepsilon_{\ell-1}}$, 
where $r$ is arbitrary and each $\varepsilon_i$ takes the value $d-1 - 2k$ for $0 \leq k \leq d-1$. The boundary condition $r_{L+1}= r_1$ is satisfied if 
$\varepsilon_1 + \ldots + \varepsilon_L = 0$.
Let ${\sf v}_\ell (= {\sf v}_\ell (s))$ be the kernel vector of ${\sf C}(r_\ell, r_{\ell+1})$ in Lemma \ref{lem:ADv} with $({\sf v}_\ell)_0 = 1$. The product-vector $\otimes_{\ell=1}^L {\sf v}_\ell (s)$, depending on $r$ and $\varepsilon_1, \ldots, \varepsilon_L$, is denoted by ${\sf v} (s ;  r , \varepsilon ) (= {\sf v} (s ; r , \varepsilon_1, \ldots, \varepsilon_L ))$. Using the first two relations in (\req(vADv)), one finds the vector ${\sf v} (s ;  r , \varepsilon )$ under the transfer matrix (\req(sft)) is decomposed as the sum 
$$
{\sf t}(s) {\sf v} (s ;  r , \varepsilon ) = a(s)^L {\sf v} (qs ;  r , \varepsilon )+ a(q^{d-1}s)^L {\sf v} (q^{-1} s ;  r , \varepsilon ).
$$
In later discussion, we need the explicit form of ${\sf v} (s ;  r , \varepsilon )$. For this purpose, we restrict the selection of $\varepsilon$ only in the form $\varepsilon_\ell = \sigma_\ell (d-1)$ with $\sigma_\ell = \pm 1$, and write $y(s| r, \sigma) (= y(s| r, \sigma_1, \ldots, \sigma_L))= {\sf v} (s ;  r , \varepsilon )$. Now the $r_\ell$ is
\be
r_\ell = r q^{(d-1)(\sigma_1 + \ldots + \sigma_{\ell-1})}, \ ~ \ {\rm with} \ \sigma_\ell = \pm 1,  \ \ \ \ \sigma_1 + \ldots + \sigma_L = 0 ,
\ele(sigma)
a condition consistent with even $L$. By (\req(vform)), the product-vector $y(s| r, \sigma)$ is expressed by
\be
y(s| r, \sigma) = \bigotimes_{\ell=1}^L \left(  \sum_{k_\ell =0}^{d-1}  r^{k_\ell}q^{k_\ell (d-1) \sum_{i=1}^\ell \sigma_i} (- s q^{\frac{-1}{2}})^{\sigma_\ell k_\ell }  \left[ {}^{d-1}_{~ k_\ell ~ }  \right]_q  {\sf e}_\ell^{k_\ell} \right),
\ele(yrs)
and satisfies the relation
$$
{\sf t}(s) y (s |r, \sigma) = a(s)^L y (qs | r, \sigma)+ a(q^{d-1}s)^L y (q^{-1} s |r, \sigma).
$$
The operator $Q_R (s)$ is a $(d-1)^L$ by $(d-1)^L$ matrix whose columns are linear combinations of vectors $y(s| r, \sigma)$. Hence follows the relation (\req(TQR)).
Similarly, using the covector-relation in (\req(vADv)), (\req(vform))  of Lemma \ref{lem:ADv}, 
one finds the product-covectors 
\be
\widehat{y}(s| r, \sigma) = \bigotimes_{\ell=1}^L \left(  \sum_{k_\ell =0}^{d-1} r^{-k_\ell} q^{-k_\ell (d-1) \sum_{i=1}^{\ell-1} \sigma_i} (s q^{\frac{-1}{2}})^{\sigma_\ell k_\ell }  {\sf e}_{k_\ell, \ell}  \right) ,
\ele(cyrs)
satisfying the relation, 
$$
\widehat{y}(s| r, \sigma) {\sf t} (s) = a(q^{d-1}s)^L  \widehat{y}(q^{-1} s| r, \sigma) + a(s)^L  \widehat{y}(qs| r, \sigma).
$$ 
Hence the relation (\req(QLT)) holds for $Q_L$-operator constructed from the covectors $\widehat{y}(s| r, \sigma)$'s. 

To construct the $Q$-operator, we need the commutative relation (\req(QLR)) about $Q_R, Q_L$. To do this, we
consider the scalar product $\widehat{y}(s' | \widetilde{r}, \widetilde{\sigma}) y (s |r, \sigma)$ of two vectors in (\req(yrs)), (\req(cyrs)), now evaluated as  
\begin{eqnarray}
\prod_{\ell=1}^L  \left(\sum_{k_\ell =0}^{d-1} (\frac{-r}{\widetilde{r}})^{k_\ell}  \left[ {}^{d-1}_{~ k_\ell ~ }  \right]_q  q^{k_\ell (d-1)(\sum_{i=1}^\ell \sigma_i -\sum_{i=1}^{\ell-1} \widetilde{\sigma}_i  )}  (\widetilde{s}q^{\frac{-1}{2}})^{\widetilde{\sigma}_\ell k_\ell} (s q^{\frac{-1}{2}})^{\sigma_\ell k_\ell}\right). \label{yfom}
\end{eqnarray}
As the argument in \cite{Bax} (9.8.23), we show the above product function is a symmetric function of $s', s$:
\be
\widehat{y}(s'| \widetilde{r}, \widetilde{\sigma}) y (s |r, \sigma) = \widehat{y}(s| \widetilde{r}, \widetilde{\sigma}) y (s' |r, \sigma).
\ele(ys's)
It is obvious that (\req(ys's)) is valid if $\widetilde{\sigma}_\ell = \sigma_\ell $ for all $\ell$. For a given $\sigma$, all values of $\widetilde{\sigma}_\ell$ allowed by (\req(sigma)) are obtained by successive interchanges of pairs $(\widetilde{\sigma}_j, \widetilde{\sigma}_{j+1})$. It suffices to show the symmetric property of the ratio function
$$
\left( \widehat{y}(s'| \widetilde{r}, ... \widetilde{\sigma}_{j+1}, \widetilde{\sigma}_j , ... ) y (s |r, \sigma) \right) / \left( \widehat{y}(s'| \widetilde{r}, ... \widetilde{\sigma}_j, \widetilde{\sigma}_{j+1}, ... ) y (s |r, \sigma) \right).
$$
Since all terms but $\ell=j, j+1$ in (\ref{yfom}) are symmetric in $\widetilde{\sigma}_j, \widetilde{\sigma}_{j+1}$, the above ratio leaves only these terms in numerator and denominator. The direct calculation on cases for $\widetilde{\sigma}_j, \widetilde{\sigma}_{j+1}, \sigma_j, \sigma_{j+1} = \pm 1$ reveals the ratio function is indeed a symmetric function of $s', s$ as the case $d=2$ in \cite{Bax}. Thus one obtains (\req(ys's)). As any column of $Q_R (s)$ is a linear combination of vectors $y (s |r, \sigma)$ and any row of $Q_L(s)$ is a linear combination of covectors $\widehat{y} (s |\widetilde{r}, \widetilde{\sigma})$, (\req(ys's)) in turn yields (\req(QLR)). We now consider the set of vectors $y (s |r, \sigma)$ formed by all possible complex numbers $r$, and all possible $\left( {}^{ ~ L}_{L/2} \right)$ values of  $\sigma = (\sigma_1, \ldots, \sigma_L)$ in (\req(sigma)); similarly for the covectors $\widehat{y} (s |r, \sigma)$. As in the six-vertex model case (\cite{Bax} page 198), conjecturally there are values of $s$ for which these vectors span the quantum vectors $\stackrel{L}{\otimes} \CZ^d$, $\stackrel{L}{\otimes} \CZ^{d *}$ respectively, so that $Q_R (s)$ and $Q_L (s)$ can be chosen non-singular. Let $s_0$ be a such value and we defined the $Q$-operator by (\req(Qdef)). Then the matrix $Q(s)$ satisfies the $TQ$-relation (\req(stQ)) .

\section{Concluding Remarks}\label{sec.F}
In this work, we first construct a $Q$-operator of XXZ chain of higher spin for the root-of-unity $q$, then another $Q$ for the general $q$, by the similar method of producing the six-vertex $Q$-operators in \cite{R06Q, Bax}, respectively. These two constructions provide different (though related) $Q$-operators of XXZ chain of higher spin, with the same $TQ$-relation. The methods may help us to understand better the $Q$-operators of solvable lattice models in general. Usually $Q$-operators in root-of-unity cases are often of particular interest, as shown in the pure ice model. In the XXZ chain of higher spin at root-of-unity $q$, the $Q$-operator with the constraint of $Q$-functional relation (\req(Qeq)) is the most important one as it encodes the symmetry of the system. Unfortunately at present I don't know any way to prove the complete set of functional relation using the $Q$-operators in this work, despite the fact that with a similar $Q$-operator, the functional relations in the spin-$\frac{1}{2}$ case have been justified in \cite{R06Q}. For higher spin cases, a better quantitative understanding about the $Q$- and fusion matrices should be essential and required for further development on the root-of-unity symmetry of the theory. A programme along this line is under consideration, and progress would be expected.

\section*{Acknowledgements}
The author is pleased to thank Professor T. Mabuchi for hospitality in November 2006 at the Department of Mathematics, Osaka University, Japan, where part of this work was carried out. 
This work is supported in part by National Science Council of Taiwan under Grant No NSC 95-2115-M-001-007.

\end{document}